\documentclass[letterpaper, 10pt, conference]{ieeeconf}
\IEEEoverridecommandlockouts               
\usepackage{graphicx}
\usepackage{epsfig,epstopdf}
\graphicspath{{../pdf/}{../jpeg/}}
\DeclareGraphicsExtensions{.pdf,.jpeg,.png}
\usepackage{amsmath,amssymb,amsfonts,mathrsfs}
\usepackage{multirow}
\usepackage{booktabs}
\usepackage{color}
\usepackage{algpseudocode}
\usepackage{tikz}
\usepackage{color}
\usepackage{cite}
\usepackage{setspace}
\usepackage{algorithm}  
\usepackage{hyperref}
\hypersetup{
colorlinks=true,
linkcolor=black,
filecolor=magenta,      
urlcolor=cyan,
pdftitle={Overleaf Example},
pdfpagemode=FullScreen,
}

\makeatletter
\def\algbackskip{\hskip-\ALG@thistlm}
\makeatother

\allowdisplaybreaks[4]
\usepackage{theorem}

\newtheorem{assumption}{Assumption}
\newcommand{\R}{\mathbb{R}}
\newcommand{\Z}{\mathbb{Z}}

\newcommand{\bma}[1]{\begin{bmatrix}#1\end{bmatrix}}

\title{\bf 
Continuous-Time System Identification and OCV Reconstruction of Li-ion Batteries via Regularized Least Squares
}

\author{
Yang Wang, Riccardo M.G. Ferrari and Michel Verhaegen
\thanks{Yang Wang, Riccardo M.G. Ferrari, and Michel Verhaegen are with the Delft Center for Systems and Control, Delft University of Technology, Delft 2628CD, Netherlands. Email: {\tt \{y.wang-40, r.ferrari, m.verhaegen\}@tudelft.nl}.} 
}
\begin{document}
\maketitle

%%%%%%%%%%%%%%%%%%%%%%%%%%%%%%%%%%%%%%%%%%%%%%%%%%%%%%%%%%%%%%%%%%%%%%%%%%%%%%%%
\begin{abstract}
Accurate identification of lithium-ion (Li-ion) battery parameters is essential for managing and predicting battery behavior. However, existing discrete-time methods hinder the estimation of physical parameters and face the fast-slow dynamics problem presented in the battery. In this paper, we developed a continuous-time approach that enables the estimation of battery parameters directly from sampled data. This method avoids discretization errors in converting continuous-time models into discrete-time ones, achieving more accurate identification. In addition, we jointly identify the open-circuit voltage (OCV) and the state of charge (SOC) relation of the battery without utilizing offline OCV tests. By modeling the OCV-SOC curve as a cubic B-spline, we achieve a high-fidelity representation of the OCV curve, facilitating its estimation. Through solving a rank and L1 regularized least squares problem, we jointly identify battery parameters and the OCV-SOC relation from the battery's dynamic data. Simulated and real-life data demonstrate the effectiveness of the developed method.
\end{abstract}
%%%%%%%%%%%%%%%%%%%%%%%%%%%%%%%%%%%%%%%%%%%%%%%%%%%%%%%%%%%%%%%%%%%%%%%%%%%%%%%%

\section{Introduction}
\label{sec: introduction}
Electric vehicles (EVs) have gained increasing attention owing to the low-carbon policy and the urgent for sustainable transportation \cite{hannan2017review}. Lithium-ion (Li-ion) batteries are now the most popular energy storage devices for EVs due to their favorable energy density, efficiency, and extended service life \cite{plett2015battery}. An accurate battery model is essential to enable a reliable EV operation because of the provision of an accurate prediction and management of the battery's performance.

Equivalent circuit models (ECMs) are battery models constructed with primary electrical elements. These models emulate battery behavior resulting from complex electrochemical processes with a simplified electrical circuit, making it affordable for onboard microprocessors of EVs \cite{yang2023improved}. Because of their decent balance between accuracy and computational costs, ECMs have been widely adopted as the battery models in EVs \cite{tian2023lithium}. Identifying ECM parameters is crucial for battery modeling as they provide critical insight for understanding and predicting battery behaviors.

Current ECM identification mainly relies on discrete-time methods \cite{wang2024concurrent,li2024online}. Though directly manageable by digital computers, these approaches are inconvenient in estimating the physical parameters of the battery. The physical processes are originally operating in continuous time, and discretization will introduce approximation error, especially when the sampling time is too large or too small compared to the time constants of the battery \cite{haverkamp1997continuous}. In addition, batteries exhibit both fast and slow dynamics due to different polarization processes. This characteristic renders the discrete-time methods more challenging in identifying the battery's parameters \cite{yang2023improved}.

Direct continuous-time identification allows for identifying a continuous-time system directly from sampled data without applying model discretization \cite{johansson1994identification,chou1999continuous}. From the continuous-time model, we can directly extract the system's physical parameters. The omission of discretization mitigates inefficiency and ill-conditioned problems in sampling a stiff system, where both fast and slow dynamics are present \cite{garnier2008direct}. These approaches are suitable for our battery identification. Continuous-time methods operate by filtering the input and output samples with a selected bank of filters. By converting the system into a linear equation of the filtered signals, these methods relate the coefficients of the resultant equation to the system parameters with algebraic transformation, which can be easily computed. Continuous-time methods benefit from omitting direct computation of derivatives, preventing the amplification of measurement noises.

The identification of batteries presents a further challenge in the problem that the open-circuit voltage (OCV) exhibits a nonlinear dependency on the state of charge (SOC) of the battery \cite{chen2006accurate}. The nonlinear characteristic renders conventional continuous-time methods not directly applicable to the ECM. Xia et al. proposed a continuous-time method for battery identification using a state variable filter and instrumental variable algorithm. In this method, offline battery tests were conducted to estimate the OCV-SOC value pairs, and the nonlinear battery system was transformed into a linear system with subtraction of the OCV from the terminal voltage of the battery \cite{xia2016accurate}. However, offline OCV tests are time-consuming and laborious. For instance, to measure one OCV-SOC value pair, the battery needs to rest for two hours to reach its equilibrium \cite{zheng2016influence}. In addition, the battery OCV tests cannot be performed once the batteries are deployed on EVs, limiting the efficacy of the test result in a condition different from laboratory settings \cite{zhang2020improved}. Shokri et al. \cite{shokri2024battery} developed a fractional-order model to identify the battery parameters of a mobile robot. The OCV-SOC curve is identified from the battery's operational data. However, this method requires the time constants of the battery to be specified a priori and thus does not reveal the actual battery parameters for providing physical interpretation.

By considering the OCV directly within the battery model, we can formulate an integrated model to simultaneously identify the battery parameters and the OCV-SOC relation. The OCV curve can be modeled with a cubic B-spline, which offers a high-fidelity representation to adjust to the piecewise nonlinearity appearance of the OCV with moderate model complexity. Integrating OCV into the dynamic model, however, leads to an interaction between the dynamic parameters and the static OCV, represented as bilinear variables in the model. We utilize the low-rank property of a structured matrix of the bilinear variables to impose the bilinearity of decision variables in a linear least squares problem. With the constructed model, the continuous-time identification can be applied using a Laguerre filter bank, which provides a more stable representation of the system than the state variable filter used in prior work \cite{xia2016accurate}. By solving a rank and L1 regularized least squares problem, we can simultaneously identify the continuous-time battery parameters and the OCV-SOC curve directly from the batteries' operational data.

\textit{Contributions:} The contributions of this paper are summarized as follows:
\begin{itemize}
    \item   We develop a continuous-time method for identifying battery parameters;
    \item We formulate the OCV-SOC curve as a cubic B-spline for an enhanced OCV estimation;
    \item The battery parameters and the OCV-SOC relation are simultaneously identified by solving a rank and L1 regularized least squares problem; 
    \item The effectiveness of the developed method is verified on a simulated battery;
    \item The developed method is adopted for real-life battery data to identify the battery parameters. 
\end{itemize}
The rest of the paper is structured as follows. Section \ref{sec: 2} introduces the equivalent circuit model for modeling Li-ion batteries. Section \ref{sec: 3} describes the proposed continuous-time method for battery identification. Section \ref{sec: 4} verifies our method with a simulated battery and real-life battery data. Section \ref{sec: 5} concludes the study and presents future work.

\section{Lithium-ion Battery Model}\label{sec: 2}
\begin{figure}[htpb!]
    \centering
    \includegraphics[width=1\linewidth]{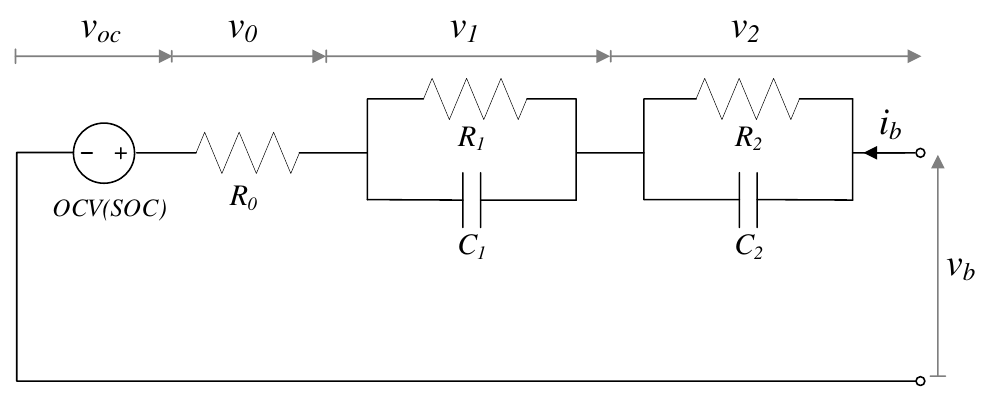}
    \caption{Second-order equivalent circuit model}
    \label{fig: ECM}
\end{figure}

Li-ion batteries can be modeled with a second-order equivalent circuit model (ECM) shown in Figure \ref{fig: ECM}.
The second-order ECM consists of an ohmic resistor $R_0$, two resistor-capacitor (RC) networks $R_1C_1$, $R_2C_2$, and an ideal voltage source $v_{oc}$. Resistor $R_0$ emulates the internal resistance of the battery, and the two RC networks represent the battery's fast and slow dynamics. Voltage source models the open-circuit voltage (OCV). Load current $i_b$ and terminal voltage $v_b$ are the model input and output, respectively. We define the charging direction as the positive direction of the current. The continuous-time state-space model of the ECM is as follows:
\begin{equation}\label{eq:state eq}
    \begin{bmatrix}
        \dot v_1(t)\\
        \dot v_2(t)\\
    \end{bmatrix}=\begin{bmatrix}
        -\frac{1}{R_1C_1} & 0\\
        0 & -\frac{1}{R_2C_2}
    \end{bmatrix}
    \begin{bmatrix}
        v_1(t)\\
        v_2(t)
    \end{bmatrix}+
    \begin{bmatrix}
        \frac{1}{C_1}\\
        \frac{1}{C_2}
    \end{bmatrix}i_b(t)
\end{equation}
\begin{equation}\label{eq:output eq}
    v_b(t)=\begin{bmatrix}
        1&1
    \end{bmatrix}\begin{bmatrix}
        v_1(t)\\
        v_2(t)
    \end{bmatrix}+
    R_0i_b(t)+v_{oc}(z(t))+n_v(t),
\end{equation}

where $v_1,v_2\in\R$ are dynamic voltages across two RC circuits, $n_v(t)\in\R$ is the measurement noise of the terminal voltage. We consider white noise $n_v$ from an i.i.d. zero-mean, $\sigma^2$-variance Gaussian process, $n_v\sim \mathcal{N}(0,\sigma^2)$. The open-circuit voltage is a nonlinear function of the state of charge (SOC), which reveals the percentage of charges contained in the battery. The SOC, denoted by $z(t)$, can be calculated via Coulomb counting as:
\begin{equation}\label{eq:soc}
    z(t)=z(t_0)+\int_{t_0}^t\frac{1}{3600C}i_b(\tau)d\tau,
\end{equation}
where $z(t_0)$ is the initial SOC at the time instant $t_0$, $C\in\R^{+}$ is the capacity of the battery in Ampere-hours (Ah), and the factor $3600$ converts Ah into Coulombs.

To identify the battery parameters, we adopt the following assumptions about the input signal and the operating conditions of the battery:

\begin{assumption}[Zero-order-hold input]\label{ass: zoh input}
The input signal is generated by a zero-order-hold machine, and the sampling and the update of the input are synchronized exactly. The output is sampled at the same time as the input.
\end{assumption}
\begin{assumption}[Constant temperature and aging]\label{ass: constant temperature and aging}
The ambient temperature of the battery is constant and does not change the battery's behavior, and the aging of the battery is considered zero during the identification.
\end{assumption}
\begin{assumption}[Constant battery parameters]\label{ass: constant parameters}
    We assume battery parameters to be constant during charging and discharging.
\end{assumption}

Assumption \ref{ass: zoh input} can be adopted for batteries employed on EVs, as the input to the battery is generated by a digital microprocessor following a ZOH scheme. Synchronization between sampling and update of the input ensures that input fed into the battery during the sampling interval takes the value of the previously sampled one. 
Assumption \ref{ass: constant temperature and aging} excludes the influence of temperature and aging of the battery on the variation of battery parameters. 
Assumption \ref{ass: constant parameters} is adopted to maintain model simplicity and efficiency. It could cause a mild error rise during extreme SOC conditions but will have a minimum effect during the typical operational range of SOC from $80\%$ to $20\%$, where battery parameters have minimum variations as reported in the literature \cite{chen2006accurate,yang2023improved}.

In this study, we aim to identify the battery parameters $R_i,C_j,i=0,1,2,j=1,2$ of the ECM \eqref{eq:state eq}\eqref{eq:output eq} and the OCV-SOC relation of the battery $v_{oc}(z(t))$ from the input signal and the output measurements of the battery.

\section{Continuous-Time Battery Identification}\label{sec: 3}
In this section, we start with deriving a regression model of the ECM required by our continuous-time identification. Then, we introduce the cubic B-spline used to model the nonlinear OCV-SOC relation. Finally, a regularized least squares problem is formulated to identify the battery parameters and the OCV-SOC relation.
\subsection{Regression model of the ECM with Laguerre filters}
To conduct continuous-time identification of the battery, we write the state-space model of the ECM \eqref{eq:state eq}\eqref{eq:output eq} into a continuous-time transfer function as:
\begin{equation}\label{eq: tf}
    G(s)=\frac{V_b(s)-V_{oc}(s)}{I_b}=\frac{b_0s^2+b_1s+b_2}{s^2+a_1s+a_2},
\end{equation}
where $s$ is the Laplace variable, and $V_b(s),I_b(s),V_{oc}(s)$ are the Laplace transforms of $v_b,i_b$ and $v_{oc}$ respectively. The coefficients of the transfer function $a_i,b_j,i=1,2,j=0,1,2$ can be transformed from the ECM parameters via algebraic relations provided at the end of this section. The transfer function \eqref{eq: tf} can be written into a linear equation as:
\begin{align}\label{eq: tf in lhs and rhs}
    (s^2+a_1s+a_2)V_b(s)=(b_0s^2+b_1s+b_2)I_b(s)+\notag\\
    (s^2+a_1s+a_2)V_{oc}(s).
\end{align}
In continuous-time identification, the differentiation of signals, represented by multiplication by $s$, is not directly available from sampled data. To tackle this issue, we convert the transfer function into a Laguerre filter basis that captures the system's dynamics. Specifically, we use the following Laguerre filter:
\begin{equation}\label{eq: laguerre}
    L_k(s)=\frac{2\nu}{s+\nu}\left(\frac{s-\nu}{s+\nu}\right)^k,
\end{equation}
where $\nu\in\R_{>0}$ is the cut-off frequency, and $k=0,1,2,$ is the order of the filter ranging from zero to the highest system order. The transfer function \eqref{eq: tf in lhs and rhs} can be converted into the following linear equation using the Laguerre filter \eqref{eq: laguerre} as:
\begin{align}\label{eq: latuerre tf}
    &(\bar{a}_0L_2(s)+\bar{a}_1L_1(s)+\bar{a}_2L_0(s))V_b(s)=\notag\\
    &\hspace{1cm}(\bar{b}_0L_2(s)+\bar{b}_1L_1(s)+\bar{b}_2L_0(s))I_b(s)+\notag\\
    &\hspace{1cm}(\bar{a}_0L_2(s)+\bar{a}_1L_1(s)+\bar{a}_2L_0(s))V_{oc}(s),
\end{align}
where the transformed parameters $\bar{a}_i,\bar{b}_i,i=0,1,2$ can be calculated from the original parameters in \eqref{eq: tf in lhs and rhs} and the cut-off frequency $\nu$ by the following relations:
\begin{align}
    \bar{a}_0&=\nu^2-a_1\nu+a_2,\
    \bar{a}_1=2\nu^2-2a_2,\label{eq: a to bar_a}\\ 
    \bar{a}_2&=\nu^2+a_1\nu+a_2,\
    \bar{b}_0=b_0\nu^2-b_1\nu+b_2,\label{eq: a to bar_a 2}\\
    \bar{b}_1&=2b_0\nu^2-2b_2,\
    \bar{b}_2=b_0\nu^2+b_1\nu+b_2. \label{eq: a to bar_a 3}
\end{align}
Detailed derivation of \eqref{eq: latuerre tf} can be found in \cite{chou1999continuous}. By applying the inverse Laplace transform to \eqref{eq: latuerre tf}, we can write the linear equation into time domain as:
\begin{align}\label{eq: time domin model laguerre}
&\bar{a}_0[L_2v_b](t)+\bar{a}_1[L_1v_b](t)+\bar{a}_2[L_0v_b](t)=\notag\\
&\hspace{0cm}\bar{b}_0[L_2i_b](t)+\bar{b}_1[L_1i_b](t)+\bar{b}_2[L_0i_b](t)+\notag\\
&\hspace{0cm}\bar{a}_0[L_2v_{oc}](z(t))+\bar{a}_1[L_1v_{oc}](z(t))+\bar{a}_2[L_0v_{oc}](z(t)).
\end{align}
where
\begin{align}
    [L_kv_b](t)&=l_k(t)*v_b(t)\\
    [L_ki_b](t)&=l_k(t)*i_b(t)\\
    [L_kv_{oc}](z(t))&=l_k(t)*v_{oc}(z(t))\, ,
\end{align}
with $*$ denoting the convolution operator and $l_k(t)$ is the impulse response of the Laguerre filter $L_k(s)$. From the time domain relation \eqref{eq: time domin model laguerre}, we can write its regression form as:
\begin{align}\label{eq: regression form}
    [L_2v_b](t)=\bma{-[L_{(1,0)}v_b](t) \hspace{-0.2cm}& [L_{(2,0)}i_b](t)}\bma{\tilde{a}\\\tilde{b}}+\notag\\
    [L_2v_{oc}](z(t))+[L_{(1,0)}v_{oc}](z(t))\tilde{a}
\end{align}
where $[L_{(i,j)}v_b]$ is the vector of filtered signals with different Laguerre bases $[[L_iv_b],[L_{i-1}v_b],\ldots, [L_jv_b]],i\geq j$, and $\tilde{a}=\bma{\tilde{a}_1 \hspace{-0.1cm}& \tilde{a}_2}^\top, \tilde{b}=\bma{\tilde{b}_0 \hspace{-0.1cm}& \tilde{b}_1 \hspace{-0.1cm}&\tilde{b}_2}^\top$ are vectors of the model parameters. $\tilde{a}_i,\tilde{b}_j$ are defined as $\tilde{a}_i:=\bar{a}_i/\bar{a}_0,\tilde{b}_j:=\bar{b}_j/\bar{b}_0$. In this formulation, we assume that $\bar{a}_0\neq 0$, which can be satisfied for a properly defined system and a suitable $\nu$, followed by \eqref{eq: a to bar_a}.

\subsection{Cubic B-spline for modeling the OCV-SOC relation}
Since the OCV $v_{oc}$ is not directly measurable from the battery and exhibits significantly different behaviors across different regions of SOC, we model it with a cubic B-spline function. A B-spline is a piecewise polynomial that enables local control of the estimated curve \cite{piegl2012nurbs,yeh2020fast}. The shape-preserving property of the B-spline suits better our OCV identification than traditional polynomial basis functions. To model the OCV as a cubic B-spline, we write $v_{oc}$ as:
\begin{equation}\label{eq: ocv spline}
    v_{oc}(z(t))=\sum_{i=1}^h\gamma_ig_i(z(t))
\end{equation}
where $g_i$ are the B-spline basis functions, $\gamma_i$ are the control points, and $h$ is the number of bases. The basis function is defined over a non-decreasing knot vector $Z:=[z_0,\ldots,z_{h+3}]$, where $z_0\leq\cdots\leq z_{h+3}$ are the positions of the splines knots. Define $p\in\Z_{\geq 0}$ the degree of B-spline. Then, the value of the basis functions can be computed recursively via the de Boor-Cox formula \cite{ma2023generalized} as:\\
for $p>0$:
\begin{align}
    g_{i,p}(z(t))=&\frac{z(t)-z_i}{z_{i+p}-z_i}g_{i,p-1}(z(t))+\notag\\
    &\frac{z_{i+p+1}-z(t)}{z_{i+p+1}-z_{i+1}}g_{i+1,p-1}(z(t));
\end{align}
for $p=1$:
\begin{equation}
    g_{i,0}(z(t))=\left 
    \{   \hspace{-0.1cm}\begin{array}{ll}
         1&  \text{if}\ z_i\leq z(t) <z_{i+1} \\
         0& \text{otherwise}
    \end{array}
    \right..
\end{equation}
In our case, $p=3$ for the cubic B-spline. A $p$-th degree B-spline has a $(p-1)$-th order continuity, with the $p$-th order derivative being a piecewise constant function. The $d$-th order derivative of the basis of B-spline can be computed as:
% The discontinuities of the $p$-th order derivative occur at the knot locations \cite{yeh2020fast}
\begin{align}\label{eq: derivative spline}
   s_{i,p}^{(d)}({z}(k))=&\frac{p}{z_{i+p}-z_i}s_{i,p-1}^{(d-1)}({z}(k))-\notag\\
    &\frac{p}{z_{i+p+1}-z_{i+1}}s_{i+1,p-1}^{(d-1)}({z}(k))
\end{align}
We will use the derivative of the cubic B-spline in a sequel for knot selection.

With the cubic B-spline \eqref{eq: ocv spline} for modeling the $v_{oc}$, we can write the Laguerre filtered version of $v_{oc}$ used in \eqref{eq: time domin model laguerre} as:
\begin{equation}
    [L_kv_{oc}](z(t))=\sum_{i=1}^{h}\gamma_i [L_kg_i](z(t)),\ k=0,1,2.
\end{equation}
This corresponds to applying the Laguerre filter to individual bases of the cubic B-spline. Then, we rewrite the regression model \eqref{eq: regression form} with cubic B-spline bases as:
\begin{align}\label{eq: regression model with spline}
    [L_2v_b](t)=\bma{-[L_{(1,0)}v_b](t) \hspace{-0.2cm}& [L_{(2,0)}i_b](t)}\bma{\tilde{a}\\\tilde{b}}+\notag\\
     [L_2g](z(t))\gamma+[L_{(1,0)}g](z(t))\tilde{a}(\gamma)
\end{align}
where $g(z(t)):=[g_1(z(t)),\ldots,g_h(z(t))]\in\R^h$ is a vector of B-spline bases evaluated at $z(t)$, and $[L_kg]$ denotes applying the Laguerre filter $L_k$ to individual elements of $g$. $\gamma:=[\gamma_1,\ldots,\gamma_h]^\top\in\R^h$ represents the vector of B-spline control points. $\tilde{a}(\gamma)\in\R^{2h}$ is a bilinear vector of $\tilde{a}$ and $\gamma$:
\begin{equation}\label{eq: a(gamma)}
    \tilde{a}(\gamma):=\tilde{a}\otimes \gamma=\bma{\tilde{a}_1\gamma\\\tilde{a}_2\gamma},
\end{equation}
where $\otimes$ is the Kronecker product.

From the formulation of the regression model \eqref{eq: regression model with spline}, we can write a data equation for the ECM as:
\begin{align}\label{eq: data eqn}
    [L_2V_b]_m=&\bma{-[L_{(1,0)}V_b]_m \hspace{-0.2cm}& [L_{(2,0)}I_b]_m \hspace{-0.2cm}& [L_2G]_m}\bma{\tilde{a}\\\tilde{b}\\\gamma}+\notag\\
    &\quad\quad\quad\quad\quad\quad\quad\quad\quad\quad\quad[L_{(1,0)}G]_m\tilde{a}(\gamma)
\end{align}
where $[L_{(1,0)}V_b]_m\in\R^{(m+1)\times 2}$ is the data matrix:
\begin{equation}
    [L_{(1,0)}V_b]_m=\bma{[L_{(1,0)}v_b](t_0)\\\vdots\\ [L_{(1,0)}v_b](t_m)},
\end{equation}
and $[L_{(2,0)}I_b]_m\in\R^{(m+1)\times 3}$, $[L_{(1,0)}G]_m\in\R^{(m+1)\times 2h}$ are  similarly defined. $t_i:=iT_s$ is the time instant of the $i$-th sample of the signal and $T_s>0$ is the sampling interval. With the formulated data equation \eqref{eq: data eqn}, we can write the following least squares problem to identify the battery parameters and the B-spline control points:
\begin{align}\label{eq: opt problem}
    \min_{\varphi,\tilde{a}(\gamma)}\ &\left\|[L_2V_b]_m-\Pi\varphi-[L_{(1,0)}G]_m\tilde{a}(\gamma)\right\|_F^2
\end{align}
where $\Pi\in\R^{(m+1)\times (2+3+h)}$ is the data matrix:
\begin{align}
    \Pi = \bma{-[L_{(1,0)}V_b]_m \hspace{-0.2cm}& [L_{(2,0)}I_b]_m \hspace{-0.2cm}& [L_2V_{oc}]_m},
\end{align}
and $\varphi=[\tilde{a}^\top,\tilde{b}^\top,\gamma^\top]^\top\in\R^{2+3+h}$ are the parameters of the transfer function and the coefficients of the cubic B-spline. Since $\tilde{a}(\gamma)$ \eqref{eq: a(gamma)} is the Kronecker product between $\tilde{a}$ and $\gamma$, \eqref{eq: opt problem} is a bilinear optimization problem that is generally expensive to solve due to nonconvexity \cite{noom2024proximal}. To impose the bilinear structure in $\tilde{a}(\gamma)$, we exploit the property that the matrix $P=\bma{M&\tilde{a}\\\gamma^\top &1}\in\R^{3\times(h+1)}$, where $M:=\tilde{a}\gamma^\top\in\R^{2\times h}$ has a rank of one:
\begin{equation}
    \text{rank}\left(\bma{M&\tilde{a}\\\gamma^\top &1}\right)=1.
\end{equation}
The matrix $M$ is formulated from the bilinear term $\tilde{a}(\gamma)$ using $\tilde{a}(\gamma)=\text{vec}(M)$, where $\tilde{a}(\gamma)$ is a row-wise vectorization of $M$. To impose the low-rank property of $P$, we regularize its nuclear norm that convexifies the rank computation. With the low-rank regularization, we formulated the bilinear optimization problem \eqref{eq: opt problem} into the following rank regularized least squares problem:
\begin{align}\label{eq: ls problem rank}
    \min_{\varphi,\tilde{a}(\gamma)}\ &\left\|[L_2V_b]_m-\Pi\varphi-[L_{(1,0)}G]_m\tilde{a}(\gamma)\right\|_F^2+\lambda_1\|P\|_*
\end{align}
where $\|\cdot\|_*$ is the nuclear norm and $\lambda_1\in\R_{\geq0}$ is the regularization coefficient.

% Constraint on feasibility of $\varphi$ imposes that the identified variables ensure positive value for the ECM resistors and capacitors. These constraints are satisfied when optimal parameters are identified due to the identifiability of the ECM \cite{xia2016accurate}. However, with regularization where the estimate might fall into an approximation, we impose these constraints to identify feasible parameters in a sequel.

\begin{figure}
    \centering
    \includegraphics[width=1\linewidth]{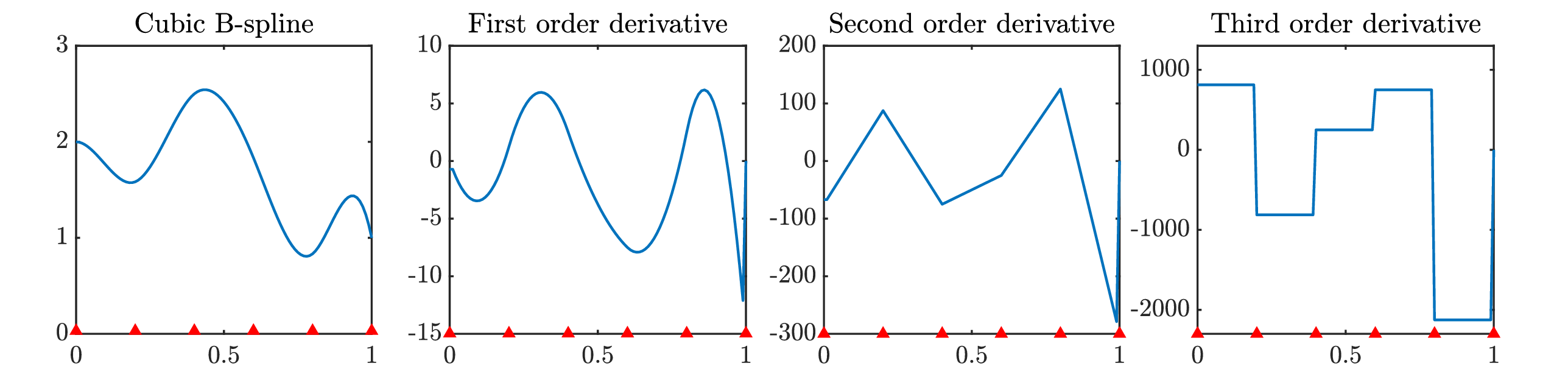}
    \caption{Cubic B-spline and its first to third-order derivatives. Knot places represented by red triangles mark the derivatives' discontinuities}
    \label{fig:spline knot place}
\end{figure}

In the regression problem \eqref{eq: opt problem}, the spline basis used for modeling OCV-SOC curve is constructed with a dense vector of knots since the structure of the OCV-SOC curve, e.g., the turning points where OCV changes significantly, are unknown a priori. However, OCV often witnesses a mild variation in a wide range of SOC \cite{chen2006accurate,zheng2016influence}. An over-dense knot vector can lead to data overfitting and sensitivity to measurement noise. To identify appropriate knot positions, we identify areas where a high variation of the B-spline is required by regularizing the sparsity of the finite difference of the third-order derivative of the B-spline. This approach utilizes the property that the third-order derivative of the cubic B-spline is a piecewise constant function, and the discontinuities occur at the knot places \cite{yeh2020fast}. This is illustrated by an example shown in Figure \ref{fig:spline knot place}. The magnitude of the B-spline's highest order discontinuities indicates the spline's variability, and places where a large magnitude exists indicate the necessity for more numbers of knots. 

To impose the sparsity, we apply an L1 regularization on the finite difference of the third-order derivative of the cubic B-spline to the least squares problem \eqref{eq: ls problem rank}. This results in the following rank and L1 regularized least squares problem:
\begin{align}\label{eq: ls rank and l1 reg}
    \min_{\varphi,\tilde{a}(\gamma)}\ &\left\|[L_2V_b]_m-\Pi\varphi-[L_{(1,0)}G]_m\tilde{a}(\gamma)\right\|_F^2+\notag\notag\\
&\quad\quad\quad\quad\quad\quad \lambda_1\|P\|_*+\lambda_2\|\mathcal{D}{G}_{m}^{(3)}\gamma\|_1
\end{align}
where $\tilde{G}_m^{(3)}$ is a matrix of bases of the third-order derivatives of the cubic B-spline in the order of sorted SOC:
\begin{equation}\label{eq: FD}
    G_{m}^{(3)}=\bma{g^{(3)}_1({z}_0) & \cdots & g^{(3)}_h({z}_0) \\g^{(3)}_1({z}_1) & \cdots & g^{(3)}_h({z}_1) \\\vdots & \ddots & \vdots \\g^{(3)}_1({z}_m) & \cdots & g^{(3)}_h({z}_m) },
\end{equation}
$g_i^{(3)}$ is the third-order derivative of $g_i$ computed by \eqref{eq: derivative spline}, and $\mathcal{D}\in\R^{m\times (m+1)}$ is a finite difference matrix:
\begin{equation}
    \mathcal{D}=\bma{1& -1 & & \\
     & \ddots& \ddots& \\
     & & 1 & -1}.
\end{equation}
In matrix \eqref{eq: FD}, sequence $\{{z}_i\}$ is a sorted SOC, where $z_i\leq z_{i+1},i=0,1,\ldots,m-1$. The sorted sequence ensures that the finite difference is computed over a monotonic SOC instead of over time, as the time series of the SOC might be non-monotonic \cite{wang2024concurrent}. The L1-norm $\|\cdot\|_1$ computes the sum of absolute values of vector elements, a convex relaxation for sparsity where the number of non-zero elements is accounted. By reducing the L1 norm, we can achieve a sparse solution in the discontinuities of the third-order derivative of the cubic B-spline, thus revealing the necessary variation of the B-spline to represent the OCV-SOC curves \cite{mustata2014receding}. 

% Since OCV is incorporated in the terminal voltage and over-regularization will lead to an underfitting of the OCV, which then affects the estimation of battery parameters, the regularization should not be overly strong.

With the parameters $\tilde{a},\tilde{b}$ identified by solving \eqref{eq: ls rank and l1 reg}, we can identify coefficients of the ECM transfer function $a_i,b_j$ in \eqref{eq: tf} by solving algebraic equations using \eqref{eq: a to bar_a}-\eqref{eq: a to bar_a 3}, and $\tilde{a}_i=\bar{a}_i/\bar{a}_0,\tilde{b}_j=\bar{b}_j/\bar{a}_0$ for $i=1,2,j=0,1,2$. Then, the physical parameters of the battery can be retrieved using the following relations:
\begin{align}
    &a_1 = \frac{1}{R_1C_1}+\frac{1}{R_2C_2},\ 
    a_2 = \frac{1}{R_1C_1R_2C_2},\  
    b_0 = R_0,\ \notag\\
    &b_1 = R_0\left(\frac{1}{R_1C_1}+\frac{1}{R_2C_2}\right)+\frac{1}{C_1}+\frac{1}{C_2},\notag\\
    &b_2 = \frac{R_0+R_1+R_2}{R_1C_1R_2C_2}
\end{align}

In the next section, we evaluate the effectiveness of the developed method for battery identification and OCV-SOC reconstruction on a simulated battery and real-life data.

\section{Numerical Experiments With Simulated and Real-Life Battery Data}\label{sec: 4}
We first validate the efficacy of the developed method by a simulated battery. Then, we apply this approach to real-life battery data from \cite{zheng2016influence}.

\subsection{Validation on a simulated battery}
The simulated battery is built with parameters from Table \ref{tab:sim param}.
\begin{table}[htbp!]
\centering
\caption{Resistors ($\Omega$) and capacitors ($F$) of the simulated battery}
\tabcolsep 9pt
\begin{tabular}{lccccc}
\hline 
Parameters & $R_0$& $R_1$& $R_2$& $C_1$& $C_2$\\ \hline
Values     & 0.06& 0.03& 0.02& 600& 5000\\ \hline
\end{tabular}
\label{tab:sim param}
\end{table}
The time constants of the simulated battery are $\tau_1=18$s and $\tau_2=100$s. The different time constants are used to validate the performance of the developed method for identifying a fast and a slow dynamic of the battery. The OCV-SOC curve is simulated with the following relation:
\begin{equation}
v_{oc}(z(t)) = 3+0.03(1.5-z(t))^{-4}+0.1\log(z(t)+0.01)\notag
\end{equation}
This function uses a different basis than polynomials to verify the generalizability of the cubic B-spline to fit various data generation functions.

\begin{figure}
    \centering
    \includegraphics[width=1\linewidth]{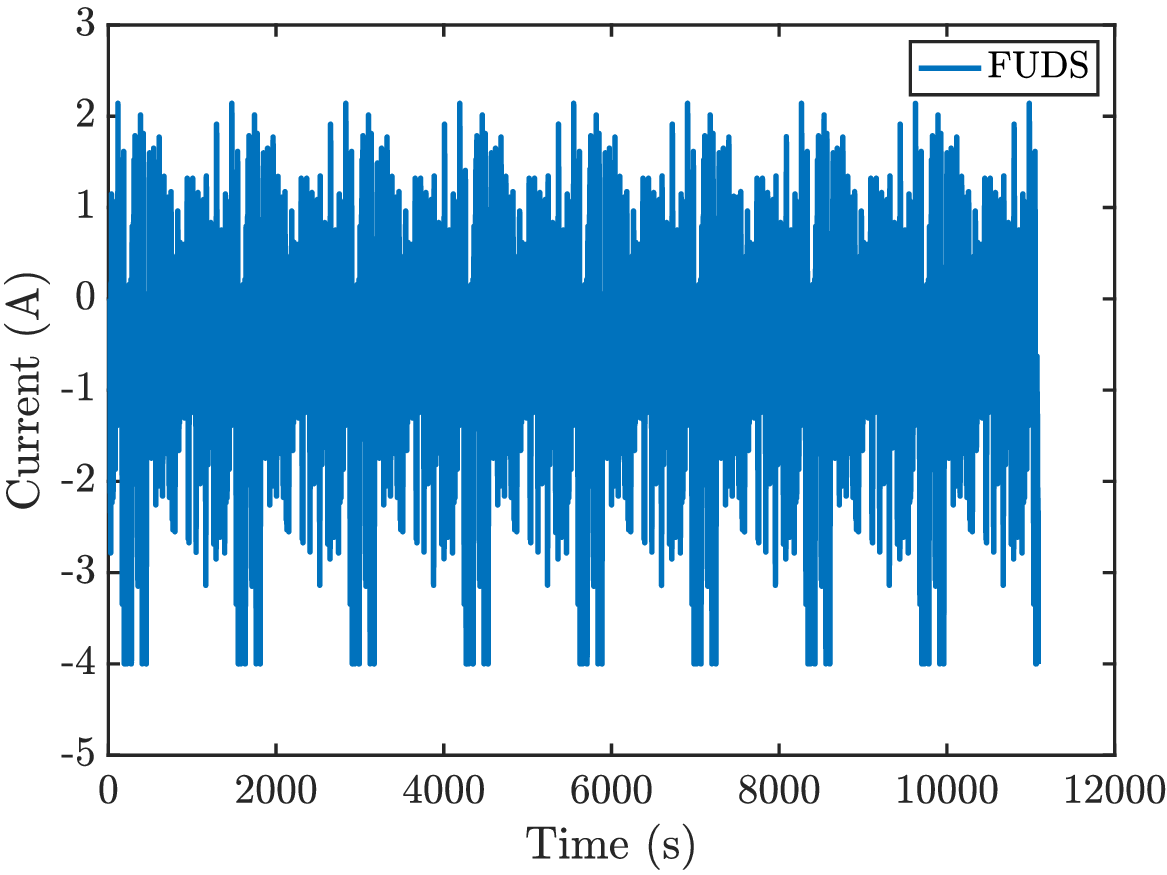}
    \caption{Federal Urban Driving Schedule Profile}
    \label{fig:fuds_profile}
\end{figure}

The input to the simulated battery is the Federal Urban Driving Schedule (FUDS) profile designed to emulate the real-life driving conditions of EVs. The input profile is shown in Figure \ref{fig:fuds_profile}. The output of the simulated battery is contaminated by white noise with a standard deviation $\sigma=10^{-4}$ to emulate measurement noise. Monte Carlo simulations with 20 different noise realizations are applied to evaluate the performance of the developed method.

We solve the rank and L1 regularized least squares \eqref{eq: ls rank and l1 reg} using data from the simulated battery to identify the battery parameters. The cut-off frequency of the Laguerre filter is $1e-3$, and the number of knots of the B-splines is $21$. The optimization problem is solved in MATLAB by \texttt{cvx} toolbox \cite{grant2014cvx} with Mosek solver \cite{mosek}. The performance of the identified model is evaluated with root mean squares error (RMSE) and variance-account-for (VAF) \cite{verhaegen2007filtering} defined by:
\begin{align}
    V_{RMSE}&= \sqrt{\frac{\sum\nolimits_{i=0}^m(v_b(t_i)-\hat{v}_b(t_i))^2}{n}},\\
    VAF&= \left(1-\frac{\text{var}(v_b(t_i)-\hat{v}_b(t_i))}{\text{var}(v_b(t_i))}\right)\times100\%,
\end{align}
where $v_b$ is the measured or simulated battery voltage and $\hat{v}_b$ is the estimated value of the model. 

To determine the regularization coefficients $\lambda_1,\lambda_2$ for the nuclear norm and L1 norm, we applied grid search on the coefficient space. For each candidate set of coefficients $\lambda_1,\lambda_2$, we solved the regularized least squares problem \eqref{eq: ls rank and l1 reg} and computed the RMSE and VAF of the identified model. We selected the coefficients that yield the minimum RMSE. The selected coefficients with minimum RMSE are $\lambda_1=2e-06$, $\lambda_2=1e-08$.

\begin{figure}
    \centering
    \includegraphics[width=1\linewidth]{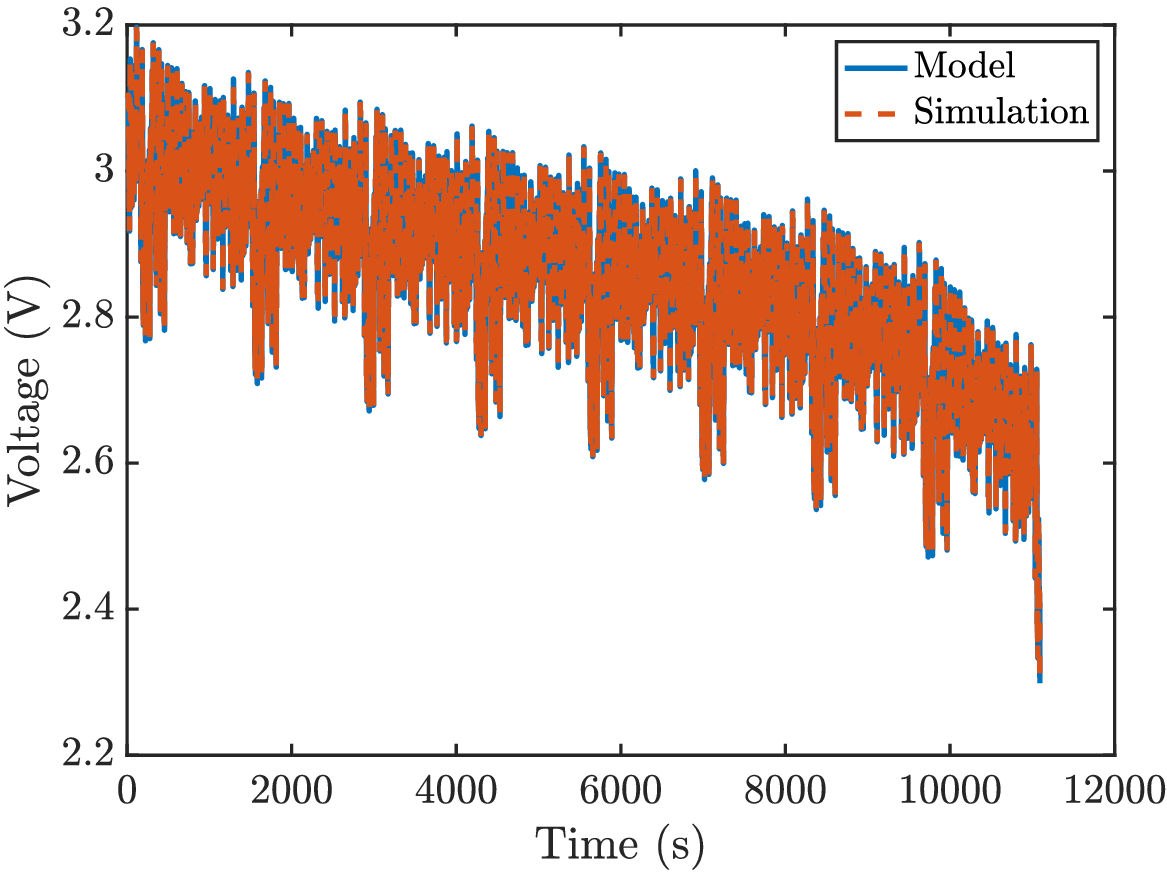}
    \caption{Model output in comparison with the simulated output}
    \label{fig:simulated_output_fitting}
\end{figure}

\begin{figure}
    \centering
    \includegraphics[width=1\linewidth]{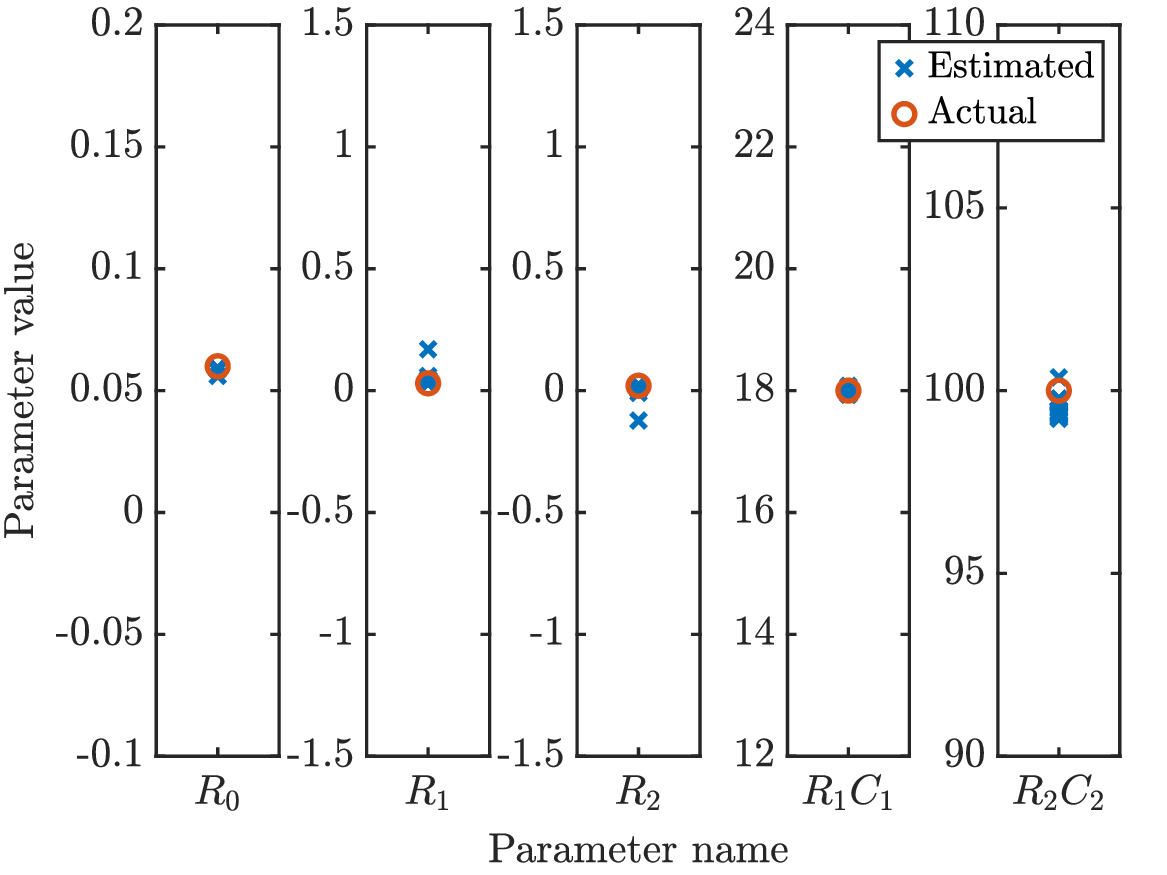}
    \caption{Estimated battery parameters in comparison to the actual values}
    \label{fig:ecm_para_est}
\end{figure}

\begin{figure}
    \centering
    \includegraphics[width=1\linewidth]{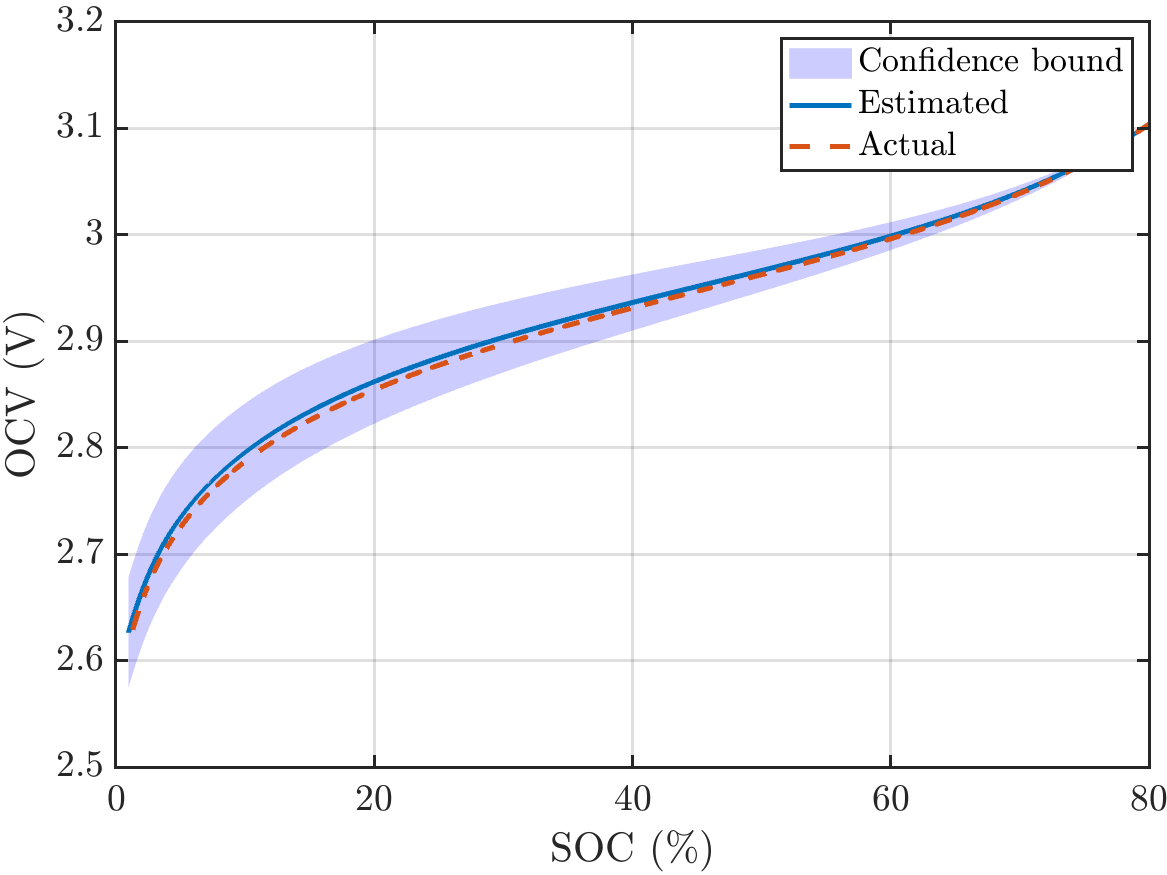}
    \caption{Estimated OCV-SOC mapping with $2\sigma$ standard deviation bounds in comparison to the actual values}
    \label{fig:ocv_est}
\end{figure}

With these coefficients, the fitting of the identified model to the simulated data is shown in Figure \ref{fig:simulated_output_fitting}. The RMSE of the model is 0.2886 mV, and the VAF is 99.74\%. The identified parameters of 20 Monte Carlo simulations are shown in Figure \ref{fig:ecm_para_est}. We can see from the figure that the estimated parameters are generally consistent with the actual values. The variation of the parameters is due to the different scales of the parameters. However, the fitting to the terminal voltage is still sufficiently accurate. The OCV-SOC identification is shown in Figure \ref{fig:ocv_est}. The estimated OCV curve is aligned with the actual value. The noise results in a mild deviation of the estimation around the simulated value. The identification is reliable for estimating the true OCV curve.


The consistency of parameter estimation and OCV reconstruction indicates the effectiveness of the developed method in simultaneously identifying the battery parameters and the OCV-SOC curve. The physical parameters with distinct time constants can also be successfully identified.

\subsection{Application to real-life battery data}
With the efficacy validated via the simulated battery, we apply the developed method to real-life battery data. The dataset we used is published by the Center for Advanced Life Cycle Engineering (CALCE) at the University of Maryland \cite{zheng2016influence}. The studied battery is discharged with the FUDS profile at a temperature of 25$^\circ$C and $80\%$ initial SOC. The OCV-SOC curve of the battery is measured under the incremental OCV test. The test results will be used as a benchmark for the identification of the OCV-SOC relation.

\begin{figure}
    \centering
    \includegraphics[width=1\linewidth]{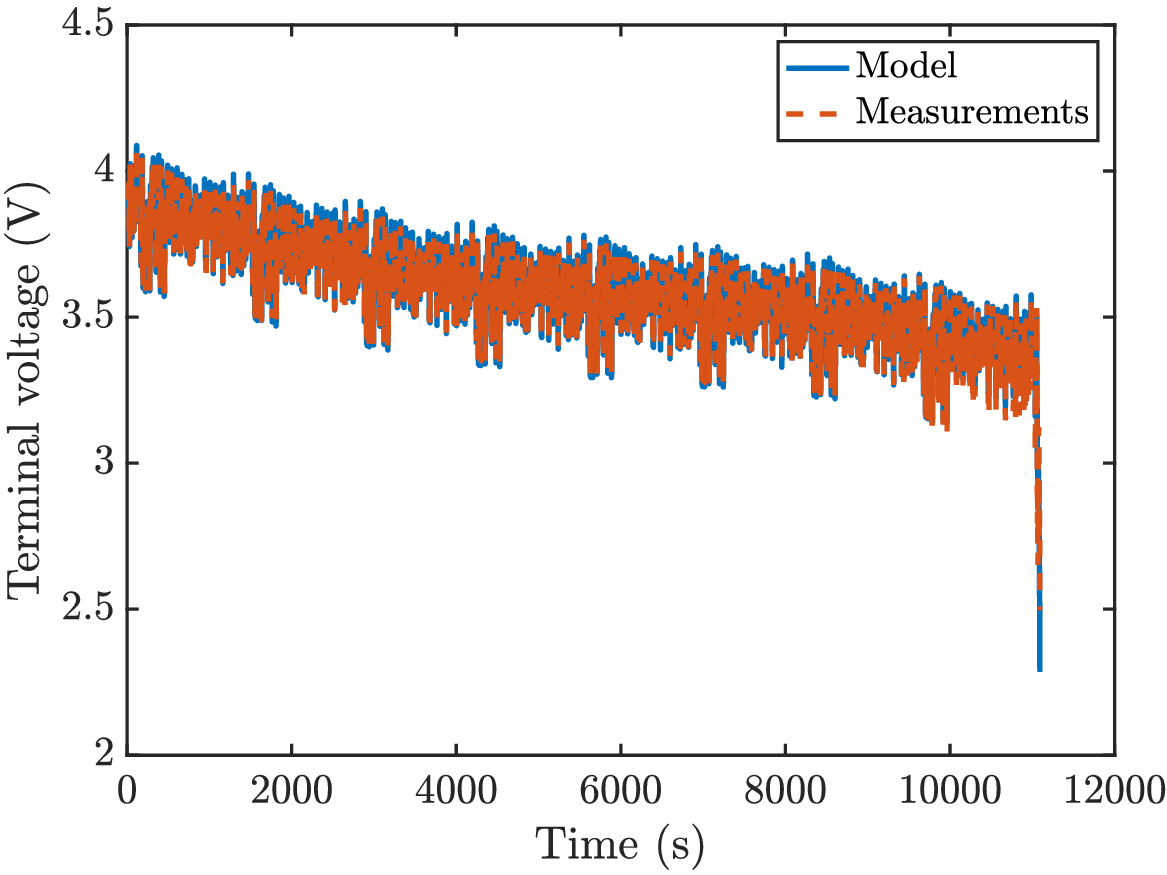}
    \caption{Model output in comparison with the measured voltage}
    \label{fig:output_fitting_real}
\end{figure}

The regularization coefficients $\lambda_1,\lambda_2$ are determined by conducting a grid search in the coefficient space as applied in the simulated battery. The selected coefficients leading to minimum RMSE are $\lambda_1 = 1.67e-06$, $\lambda_2 = 1.00e-08$. The fitting of the terminal voltage of the identified model is shown in Figure \ref{fig:output_fitting_real}. The RMSE of the voltage fit is 0.0156V, and the VAF of the model is 99.3522\%. The identified battery parameters are shown in Table \ref{tab:est real param}.  The time constants of the tested battery are $1.55$s and $30.94$s. This result indicates the fast-slow dynamics of the battery.

\begin{figure}
    \centering
    \includegraphics[width=1\linewidth]{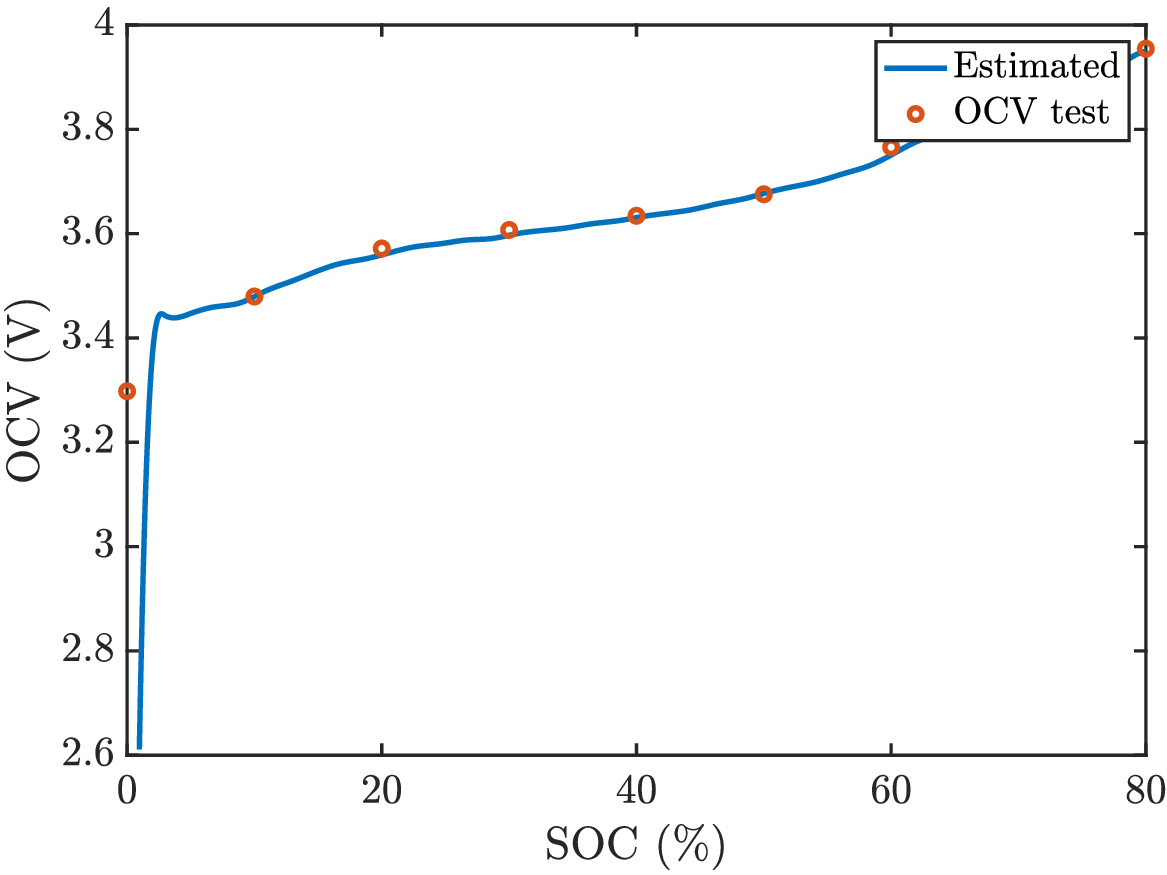}
    \caption{Estimated OCV-SOC mapping in comparison with the results of the OCV test}
    \label{fig:ocv_est_real}
\end{figure}

The reconstructed OCV-SOC curve is pictured in Figure \ref{fig:ocv_est_real}. The figure shows that the identified OCV is aligned with the result of the OCV test. Moreover, the estimation provides more detailed information on the OCV curve during intervals between two samples of the OCV tests, especially in the range from $60\%$ to $70\%$. The drastic decrease of the OCV within $0\%$ to $10\%$ SOC indicates a large voltage drop of the battery when approaching a fully discharged state. This information cannot be attained from the battery OCV tests but is feasible by direct estimation of the OCV-SOC relation. The validated result illustrates the advantages of directly identifying the OCV-SOC relation from operational data over conducting OCV tests.

% \begin{table}[htbp!]
% \centering
% \caption{Identified resistors ($\Omega$) and capacitors ($F$) of the tested battery}
% \label{tab:est real param}
% \begin{tabular}{llll}
% \hline
% Parameters & R0  & R1  & R2  \\ \hline
% Values     & 0.1 & 0.2 & 0.3 \\ \hline
% Parameters & C1  & C2  &     \\ \hline
% Values     & 100 & 200 &     \\ \hline
% \end{tabular}
% \end{table}

\begin{table}[htbp!]
\centering
\caption{Identified resistors [$\Omega$] and capacitors [$F$] values  of the tested battery}
\label{tab:est real param}
\tabcolsep 3pt
\begin{tabular}{lccccc}
\hline 
Parameters & $R_0$& $R_1$& $R_2$& $C_1$& $C_2$\\ \hline
Values     & 0.0648& 0.0105& 0.0158& 147.9061& 1958.39\\ \hline
\end{tabular}
\end{table}

\section{Conclusion}\label{sec: 5}
We developed a continuous-time method for identifying the battery parameters and the OCV-SOC relation directly from the battery's operational data. Laguerre filters were utilized to achieve continuous-time identification. By modeling the OCV-SOC relation with a cubic B-spline, we realize a high-fidelity representation of the piecewise nonlinearity appearance of the OCV in its identification. The battery parameters and the OCV curve are simultaneously identified by solving a rank and L1 regularized least squares problem. Simulation experiments and real-life data validate the effectiveness of the developed method. Future work will study optimal methods for determining regularization coefficients and a recursive approach to track variable battery parameters. 

\bibliographystyle{ieeetr}
\bibliography{library} 
\end{document}